\documentclass[usenatbib]{mn2e}
\usepackage{graphicx}

\begin{document}

\title[Enhanced mergers] 
{Enhanced mergers of galaxies in low-redshift clusters}

\author[C. Moss]
{C.~Moss \\ 
Astrophysics Research Institute, Liverpool John Moores
University, Birkenhead CH41 1LD }

\maketitle

\begin{abstract}
An ensemble cluster has been formed from a data set comprising a complete
magnitude-limited sample of 680 giant galaxies 
($M_{B}^{0}$ \raisebox{-1ex}{$\stackrel{\textstyle <}{\sim}$} $-19$)
in 8 low-redshift clusters, normalised by the velocity dispersions and
virial radii for the early-type cluster populations.  Distinct galaxy
populations have been identified, including an infall population.  A
majority (50--70\% or greater) of the infall population are found to
be in interacting or merging systems characterised by slow
gravitational encounters. The observed enhancement
of galaxy--galaxy encounters in the infall population compared to the
field can be explained by gravitational shocking.  It is shown that disc
galaxy mergers in the infall population integrated over the estimated
lifetime of the cluster ($\sim$ 10 Gyr) can readily account for the
present cluster S0 population.
\end{abstract}


\section{Introduction}

Although the transformation of luminous cluster disc galaxy populations from
mainly spirals to predominantly S0s between intermediate redshifts ($z
\sim 0.5$) and the present is well established (e.g. Dressler 1980;
Dressler et al. 1997), there is as yet no agreed mechanism for this
transformation. One of the earliest explanations proposed was
galaxy--galaxy interactions and mergers (e.g. Lavery \& Henry 1988;
Lavery, Pierce \& McClure 1992). This explanation is attractive both
on account of the observed abundance of interacting/merging systems in
intermediate redshift clusters (e.g. Dressler et al. 1994a, 1994b) and
because the Hubble sequence is readily explainable as a sequence of
decreasing merger damage (Schweizer 2000). Nevertheless this
explanation was resisted because it was considered that slow
galaxy--galaxy interactions and mergers are likely to be
rare in virialised clusters (e.g. Ostriker 1980; Makino \& Hut 1997;
Ghigna et al. 1998).  Although many other explanations have been
attempted, including ram-pressure stripping of the cold interstellar
gas of spirals by the hot ionised intracluster medium (e.g. Quilis,
Moore \& Bower 2000), galaxy `harassment', the impulsive heating of a
galaxy disc by high-speed encounters (e.g. Moore et al. 1996, 1999),
and `strangulation', the stripping of an (hypothesised) hot halo gas from
spirals (e.g.  Larson et al. 1980; Bekki, Couch \& Shioya 2002), these
explanations are not without difficulties (for a recent discussion,
see Mihos 2004).  Alternatively, attempts have been made to remove the
problem entirely from clusters by proposing preprocessing of disc
galaxies to earlier type systems in groups in the field (e.g. Balogh
et al. 2004).

However, as remarked by Mihos (2004), neglect of slow encounters has
been premature. Numerical simulations which track galaxy halos in a
rich cluster in an $\Omega_{m} = 1$ cosmology show that between
$z = 0.5$ and $z=0$, although no mergers occur in the central virialised region
of the cluster, in the outskirts the merger rate is 5--9\% (see Ghigna
et al. 1998). Gnedin (2003) has similarly studied the
interaction and merger rates in clusters under different cosmologies.
Over the lifetime of a cluster, the distribution of encounter velocities
shows a large tail, but a significant fraction of encounters, largely
those in the cluster periphery or those occurring at higher redshift before
the cluster has fully collapsed, have relatively low velocity.  For example,
for an $\Omega_{m} = 0.4$ $\Omega_{\lambda} = 0.6$ cosmology, some 42\%
of encounters have a velocity $\le$ 550 km ${\rm s}^{-1}$.

As Mihos (2004) has noted, galaxy clusters form not by accreting individual
galaxies randomly from the field environment, but rather through the 
infall of less massive groups falling in along the filaments that make 
up the `cosmic web'. Such infalling
groups provide sites with much lower velocity dispersions than that of
the cluster, thus permitting strong, slow encounters more normally
associated with the field. For low-redshift clusters, such infall
continues at the present (e.g. Moss \& Dickens 1977; Tully \& Shaya
1984; Sodre et al. 1989; Colless \& Dunn 1996; Biviano et al. 1997;
Rines et al. 2003). 

Besides the theoretical considerations outlined above, there is also
accumulating observational evidence of the potential importance of
galaxy mergers for morphological transformation of cluster disc galaxies,
both at low and intermediate-redshift (e.g. Dressler et al. 1997;
Moss, Whittle \& Pesce 1998; Moss \& Whittle
2000, 2005; Koopmann \& Kenney 2004; Sato \& Martin 2006).

In the present paper the infall population of nearby clusters is
studied using data from an extensive morphological and star formation
survey of 8 low-redshift clusters (Moss \& Whittle 2000, 2005). These
data are used to construct an ensemble cluster from which general
properties of all the clusters can be deduced.  It is shown that a
very high fraction ($\sim$ 50--70\%) of the infall population are
either interacting or possible mergers.  This interaction/merger rate
is too high to be accounted for simply on the basis of infalling
groups typical of the field, but implies that the accretion
process of the cluster enhances the galaxy--galaxy interaction/merger
rate.  It is suggested that galaxy shocking (Struck 2005) can provide
a mechanism to enhance the interaction/merger rate, and that 
galaxy--galaxy interactions and mergers can provide an explanation for
the transformation of most luminous cluster spirals to S0s over the past
$\sim$ 10 Gyr.

The paper is organised as follows.  In section 2, the survey data are
summarised and construction of an ensemble cluster is discussed.  In
section 3, various sub-populations of the ensemble cluster are
identified, including that of the infall population. Member galaxies
of the infall population are considered in detail and an estimate of
the interaction/merger rate for this population is given. Discussion
is given in section 4. It is shown that there is an enhancement of
galaxy mergers in the infall population in comparison to the field; an
explanation (gravitational shocking) is proposed for this enhancement;
and finally it is shown that it is entirely plausible that the
majority of the current S0 cluster population have resulted from
galaxy mergers in the past $\sim$ 10 Gyr.  Conclusions are given in
section 5.
  
\section{H$\alpha$ survey}

\subsection{Survey galaxy sample and detected emission}

Using the 61/94-cm Burrell Schmidt telescope on Kitt Peak, an
objective prism survey was undertaken of combined H$\alpha$ + [NII]
emission in CGCG galaxies (Zwicky et al. 1960--68) in the low-redshift
clusters, Abell 262, 347, 400, 426, 569, 779, 1367 and 1656. The
surveyed galaxies were all those which lie within a radial distance, $r
\le 1.5$ Abell radii (${\rm r}_{\rm A}$; cf. Abell 1958) of the cluster
centres\footnote{${\rm r}_{\rm A} = 2.14\:{\rm Mpc}$, assuming a value
of the Hubble constant, $H_{0} = 70\:{\rm km}\:{\rm s}^{-1}\:{\rm
Mpc}^{-1}$.  This value is assumed throughout the paper.}  ; and in
addition 79 galaxies, mainly in Abell 1367, which lie in a
supercluster field region (viz. $1.5{\rm r}_{\rm A} < r \le 2.6{\rm
r}_{\rm A}$).  The full survey sample (including separate components
of double galaxy systems) comprises 843 galaxies.  The sample is
magnitude-limited; redshifts for the galaxies are 96\% complete.
Further details of the survey, including discussions of galaxy
classification and survey completeness, can be found in previous
papers (cf. Moss, Whittle \& Irwin 1988; Moss \& Whittle 1993; Moss,
Whittle \& Pesce 1998; Moss \& Whittle 2000, 2005;
hereafter, papers I--V respectively).

Emission was detected in 5\% of early-type (E--S0/a) and 41\% of
late-type (Sa + later) galaxies respectively.  The survey
distinguished between {\it compact} and {\it diffuse} emission.  The
former has been identified as predominantly due to circumnuclear
starburst emission, while the latter is characteristic of more normal
disc star formation.  The compact HII emission is associated with a
bar in the galaxy (significance level $3.1\sigma$); with a
tidally-disturbed morphology (significance level $8.7\sigma$); and
with the presence of a nearby galaxy companion (significance level
$3.1\sigma$).  It is found that compact emission, particularly that
associated with a tidally-disturbed galaxy morphology, is enhanced in
clusters as compared to the field (cf. papers IV and V).

\subsection{Ensemble cluster}
\label{syncl}

For the present study, data for the 8 clusters have been combined into
a single ensemble cluster. This procedure necessarily erases
structural details for the individual clusters.  However, as will be
shown, important properties of all the clusters can be found from the
combined data which would otherwise not be revealed by data from a
single cluster alone.

The sample galaxies were combined into an ensemble cluster as follows:
velocities were normalised by the cluster velocity dispersion,
$\sigma$, and scaled to the cluster mean, $\bar{v}$; radial distances
from the cluster centres (Abell 1958) were normalised by the virial
radius, ${r}_{vir}$.

As will be shown in what follows, the cluster late--type galaxy
population (types Sa + later) has an infalling component with higher
velocity dispersion than the early-type cluster galaxies, as well as a
component whose velocity distribution is asymmetric with respect to
the cluster mean.  In contrast, the early-type cluster population has
a Gaussian distribution expected for virially relaxed galaxies in the
cluster.  A recent study of 59 clusters in the ESO Nearby Abell
Cluster Survey has also confirmed that deviations from a Gaussian
velocity distribution for early-type galaxies in the ensemble
of these clusters are very small (cf. Katgert, Biviano \& Mazure
2004).  Moreover the early-type galaxies are generally considered to
be the oldest cluster population and therefore most likely to be in
dynamical equilibrium with the cluster potential.  For these reasons,
cluster mean (heliocentric) velocities and velocities dispersions have
been determined from the early-type galaxies alone.  For each cluster,
$\bar{v}$,$\sigma$ were determined using biweight estimators of
central location and scale (cf. Beers, Flynn \& Gebhardt 1990;
Teague, Carter \& Gray 1990).  The
initial galaxy sample was all E,S0,S0/a galaxies with $r \le {\rm
r}_{\rm A}$ and $\left| v - \bar{v} \right| \le 4\sigma$, where
initial values for $\bar{v}$ and $\sigma$ were taken from Struble \&
Rood (1991). Solutions for $\bar{v}$, $\sigma$ were determined using 3
iterations and are listed in Table \ref{vradii}.

Values of the virial radii for the individual clusters were estimated
using the relation $r_{vir} \simeq 3.5\sigma(1+z)^{-1.5}$ (see Lewis
et al.  2002), where $r_{vir}$ is in Mpc for $\sigma$ in units of 1000
km ${\rm s}^{-1}$. The derivation of this relation assumes spherical
symmetry, and that the galaxy distribution follows the mass
distribution for the cluster (see  Girardi et al. 1998).  The virial
radii thus determined are listed in Table \ref{vradii}. Abell 569 is a
double cluster (north and south sub-cluster centres (J2000): 7$^{\rm
h}$ 10\fm6 +50$^{\circ}$ 07$^{\prime}$ and 7$^{\rm h}$ 9\fm1
+48$^{\circ}$ 37$^{\prime}$ respectively); the radial distance from
the cluster centre for an individual galaxy was taken as its distance
from the nearest sub-cluster centre.

Using the values of $\bar{v}$,~$\sigma$ and ${\rm r}_{vir}$ given in
Table \ref{vradii}, the sample galaxies for the 8 clusters were
combined into a single ensemble cluster sample. Of the 843 galaxies in
the original sample, 39 were excluded which are components of double
systems with an estimated magnitude fainter than the survey limit
($m_{p} = 15.7$); 90 foreground/background galaxies (
$\left| v - \bar{v} \right| > 4\sigma$) were also excluded. 
Of the remaining 714
galaxies, 17 galaxies have no measured velocity and 17 are untyped.
The remaining 680 galaxies (552 with $r \le r_{vir}$) comprise the
galaxy sample for the analysis which follows.

\begin{table}
\centering
  \caption{\label{vradii} Cluster virial radii}
\vspace{\baselineskip}
\begin{tabular}[h]{lccccc} \hline
\multicolumn{1}{l}{\rule[-2mm]{0mm}{4mm} \hspace*{-0.5em} Cluster} &
\multicolumn{1}{c}{$\bar{v}$}  &
\multicolumn{1}{c}{$\sigma$} &
\multicolumn{1}{c}{$n$} &  
\multicolumn{2}{c}{$r_{vir}$} \\ \cline{5-6}
\multicolumn{1}{c}{\rule[-2mm]{0mm}{4mm} \hspace*{-0.5em} (E--S0/a only)} &
\multicolumn{1}{c}{(km ${\rm s}^{-1}$)} &
\multicolumn{1}{c}{(km ${\rm s}^{-1}$)} &&
\multicolumn{1}{c}{(Mpc)} &
\multicolumn{1}{c}{(${\rm r}_{\rm A}$)} \\ \hline
Abell  262 & 4812 &  537 &  38 & 1.83 & 0.86 \\ 
Abell  347 & 5582 &  550 &  14 & 1.87 & 0.87 \\
Abell  400 & 6771 &  392 &  10 & 1.32 & 0.62 \\
Abell  426 & 5161 & 1076 &  55 & 3.67 & 1.72 \\
Abell  569 & 5868 &  417 &  26 & 1.42 & 0.66 \\
Abell  779 & 6991 &  290 &  11 & 0.98 & 0.46 \\
Abell 1367 & 6542 &  762 &  62 & 2.58 & 1.21 \\
Abell 1656 & 6944 &  890 & 131 & 3.01 & 1.41 \\ \hline
\end{tabular}
\end{table}

\section{Galaxy populations}
\label{gpops}

\begin{figure}
\centering
\includegraphics[width=0.368\textwidth,angle=-90,bb=66 71 540 676]{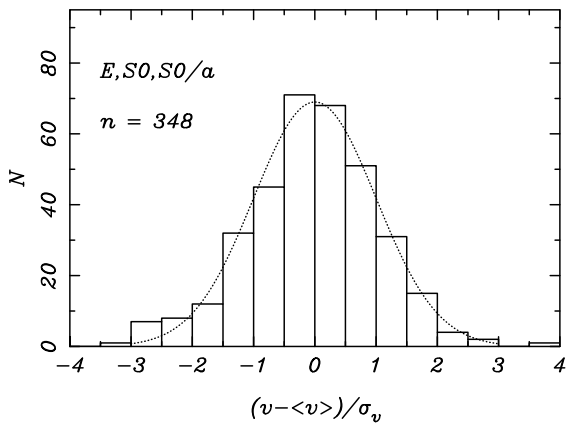}
\caption{\label{vde} Normalised velocity distribution scaled to the 
cluster mean velocity, for E, S0, S0/a galaxies in the ensemble
cluster ($r \le r_{vir}$). The dotted line is a normalised
Gaussian fit to the data ($\left| v - \bar{v} \right| \le 3\sigma$)
with mean of 0 and dispersion of 1. }
\end{figure}

\begin{figure}
\centering
\includegraphics[width=0.368\textwidth,angle=-90,bb=71 91 519 677]{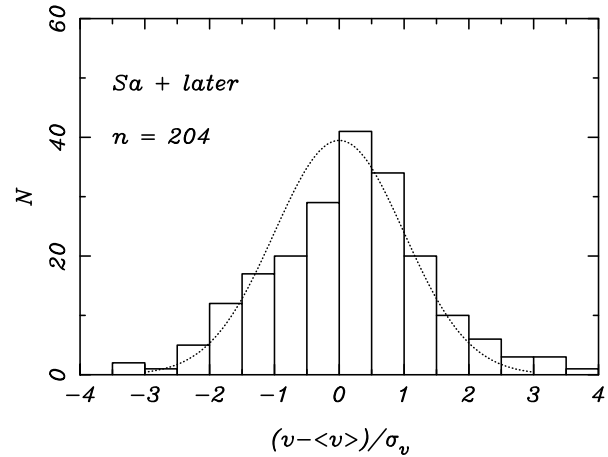}
\caption{\label{vds} As Figure \ref{vde} for  
Sa + later galaxies.}
\end{figure}

The distribution of normalised velocities scaled to the cluster mean,
$(v-\bar{v})/\sigma$, for E,S0,S0/a galaxies in the ensemble cluster
($r \le r_{vir}$) is shown in Figure \ref{vde}. Also shown in
the figure is a normalised Gaussian fit to the data ($\left| v -
\bar{v} \right| \le 3\sigma$) with mean of 0 and dispersion of 1. As
is seen, the data are well fitted by a Gaussian (K--S test,
significance level = 0.53), consistent with the early-type cluster
galaxy population being virially relaxed.

In Figure \ref{vds}, the same velocity distribution for cluster
galaxies ($r \le r_{vir}$) of types Sa + later is shown. This
distribution shows some positive asymmetry with respect to the
normalised Gaussian. In order to better understand this distribution, it is 
useful to consider sub-populations of the cluster galaxy sample.

In previous work it was shown that galaxies of types Sa + later
with disturbed stellar populations
are more frequently found in clusters as compared to the
field; moreover, such disturbed galaxies with compact HII emission
have few counterparts in the surrounding supercluster field, and
substantially account for the observed enhancement of galaxies with
compact emission in clusters (cf. Papers IV and V). Accordingly it is
convenient to begin the analysis of cluster galaxy sub-populations by
distinguishing between galaxies with disturbed and undisturbed
morphologies respectively.

In section \ref{cip} below, it will be shown that the galaxies with a
disturbed morphology, which are predominantly of types Sa + later,
most likely comprise an infall population to the clusters.  In contrast,
galaxies of types Sa + later with an undisturbed morphology are
expected to be more typical of the outer cluster regions or the field
(cf. section \ref{chp}). The observed asymmetry of the velocity
distribution for Sa + later is due to HII emission-line galaxies (ELGs)
with an undisturbed morphology, which are likely to be members of
galaxy groups accreting onto the clusters (see section \ref{pfg}).

It is to be noted that those few ELGs with known AGN or LINER emission
($n = 10$; $\sim$ 10\% of ELGs) have been omitted from the analysis,
since the main interest of the present study is in the effect of
environment on galaxy star formation.  In fact, inclusion of these
galaxies in the ELG samples does not significantly change the results
obtained below.

\subsection{Cluster infall population} 
\label{cip}

\begin{figure}
\centering
\includegraphics[width=0.613\textwidth,angle=-90,bb=83 231 484 545]{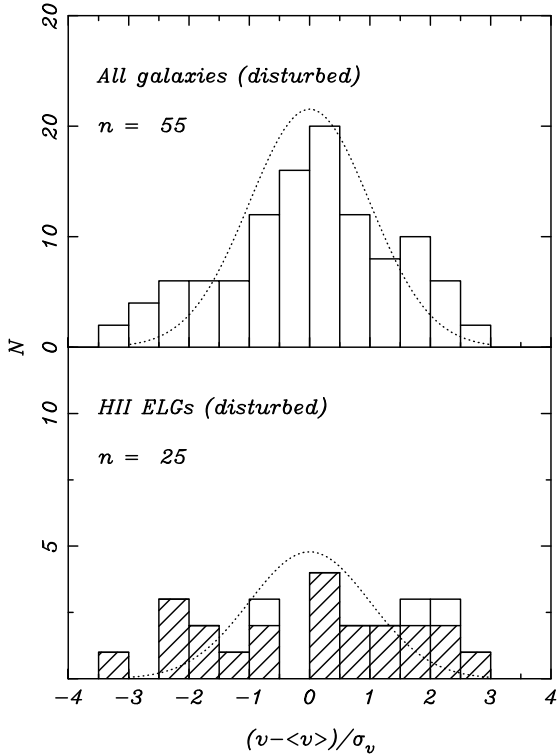}
\caption{\label{vdd} As Figure \ref{vde} for (upper panel) all 
galaxies with a disturbed morphology and (lower panel) for HII ELGs with
a disturbed morphology. The hatched area of lower histogram shows the 
distribution of galaxies with compact emission.}
\end{figure}

\begin{figure}
\centering
\includegraphics[width=0.35\textwidth,angle=-90,bb=61 92 536 657]{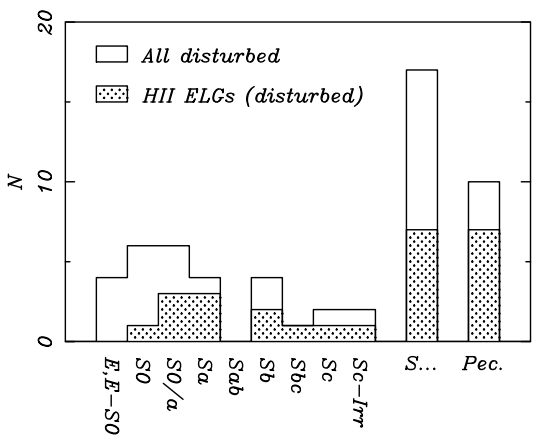}
\caption{\label{ttd} Distribution of galaxy types for all disturbed galaxies
and (shaded histogram) HII ELGs with a disturbed morphology.}
\end{figure}

\begin{figure}
\centering
\includegraphics[width=0.613\textwidth,angle=-90,bb=83 231 484 545]{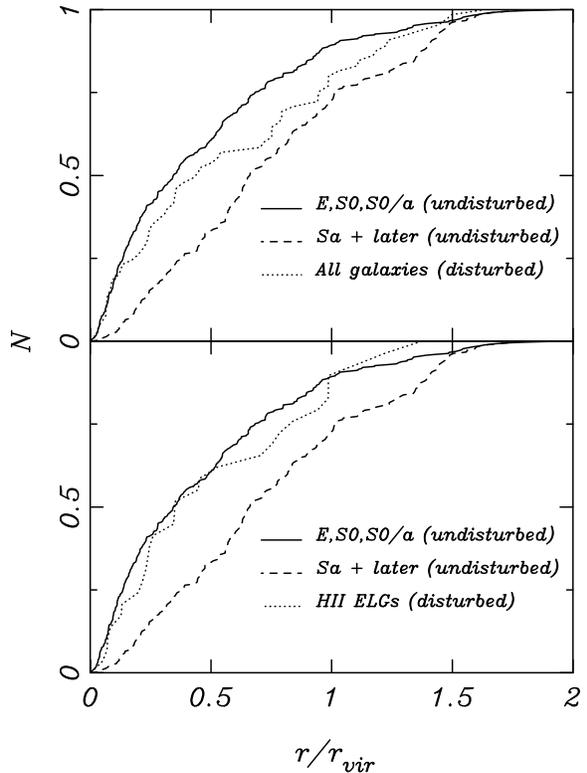}
\caption{\label{cdd} Cumulative distributions with $r_{vir}$ for 
galaxy types (excluding galaxies with a disturbed morphology)
E,S0,S0/a (solid line, $n = 354$) and Sa + later (dashed line, $n =
216$) in the ensemble cluster.  Also shown are the cumulative
distributions for (upper panel) all galaxies with a disturbed morphology
(dotted line, $n = 65$), and (lower panel) HII ELGs with a
disturbed morphology (dotted line, $n = 29$).  All samples exclude
galaxies with $\left| v - \bar{v} \right| > 4\sigma$.}
\end{figure}

\begin{figure}
\centering
\includegraphics[width=0.613\textwidth,angle=-90,bb=83 231 484 545]{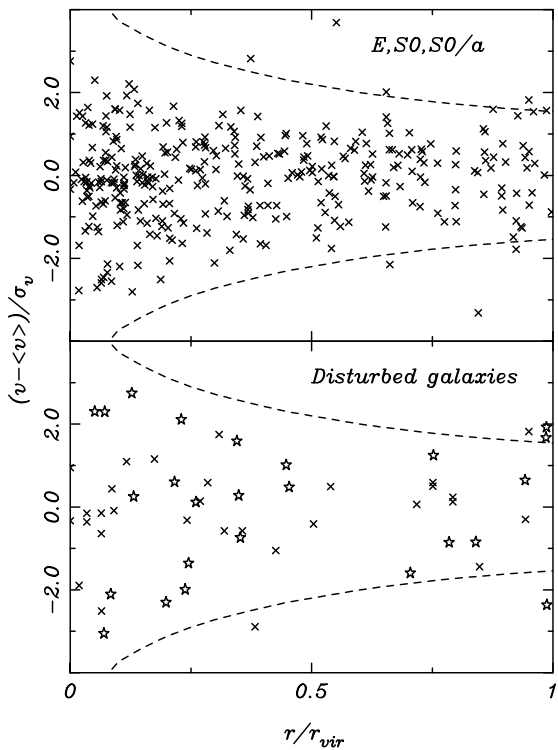}
\caption{\label{vrdd} Distribution in $r$--$v$ plane of (upper panel) E,S0,S0/a
galaxies, and (lower panel) all disturbed galaxies.  In the lower panel, 
HII ELGs are shown as stars. The dashed curves are caustics which show the
maximum line of sight velocity for a radial infall population.}
\end{figure}

In Figure \ref{vdd} (upper panel) the normalised velocity distribution
of all galaxies with a disturbed morphology in the ensemble cluster
($r \le r_{vir}$) is shown. Classification of galaxies as disturbed or
undisturbed has been taken from Papers III--V. The mean and dispersion
(biweight estimators) for the velocity distribution are $0.05 \pm
0.19$ and 1.42 respectively. The mean is in agreement with that for
the early-type (E, S0, S0/a) galaxies; however the dispersion is
significantly greater than that expected for a sub-sample of the
virialised cluster population (F test, significance level = $5 \times
10^{-4}$). In fact, the dispersion ($\sigma_{v} = 1.42$, with 90\%
confidence limits, 1.17, 1.56 estimated from bootstrap resampling)
is in good agreement with the expected value of $\sqrt{2}$ for an
infall population.

In the lower panel of the figure is shown the corresponding velocity
distribution for HII ELGs with a disturbed morphology. The ELGs
classified as having compact emission are shown in the hatched
histogram.  The mean and dispersion for the full HII ELG sample are
$0.14 \pm 0.35$ and 1.75 (90\% confidence limits: 1.38, 1.93). Again,
the mean is in accord with the cluster mean; however the dispersion is
significantly greater than that expected for a sub-sample of the
virialised cluster population (F test, significance level = $2 \times
10^{-5}$), and even higher than for the sample of all disturbed
galaxies.

The samples of all disturbed galaxies, and HII ELGs with a disturbed
morphology are predominantly of types Sa + later (71\% and 85\%
respectively, see Figure \ref{ttd}).  The velocity distribution for
disturbed galaxies, restricted to types Sa + later, is intermediate
between those for the two other samples with mean and dispersion of
$-0.01 \pm 0.25$ and 1.55 respectively ($n = 39$).

In the upper and lower panels of Figure \ref{cdd} are shown the
cumulative spatial distributions with $r_{vir}$ of the samples of all
disturbed galaxies, and HII ELGs with a disturbed morphology
respectively.  In each case the sample is compared to the cumulative
distributions for (undisturbed) galaxies of types E,S0,S0/a and Sa +
later.  It is seen that the sample of all disturbed galaxies has a
more concentrated distribution than that for undisturbed galaxies of
types Sa + later.  Remarkably, the distribution of the sample of HII ELGs
with disturbed morphology is similar to that of the early-type
population (K--S test, significance level = 0.38) and more
concentrated than the (undisturbed) Sa + later galaxies (K--S test,
significance level = $7 \times 10^{-3}$) despite the fact that this
ELG sample is predominantly comprised of types Sa +
later (see Figure \ref{ttd}).

The simplest explanation for the higher velocity dispersion observed
for all disturbed galaxies is that the disturbed galaxies are an
infall population whose velocities have not yet been virialised. In
Figure \ref{vrdd} (lower panel) the distribution of disturbed galaxies
in the $r$--$v$ plane is shown, as compared to the corresponding
distribution for E,S0,S0/a galaxies (upper panel).  In the figure, HII
ELGs with a disturbed morphology are shown as stars in the lower
panel. In both panels, the dashed curves are caustics which are a
measure of the cluster gravitational potential (Diaferio 1999); they
show the maximum line of sight velocity for a radial infall
population, as a function of distance from the cluster centre.
These caustics were calculated using a standard NFW mass
density profile for the E,S0,S0/a population with an assumed value
of the concentration parameter, $c = 5.56$ (see Navarro, Frenk \&
White 1997; Biviano \& Girardi 2003).

For a virially-relaxed population it is expected that points in the
figure will cluster towards the axis, \mbox{$(v - <v>)/\sigma_{v} =
0$;} in contrast, for a radial infall population with a similar mass
density profile the density of points in the figure is expected to
increase towards the caustics, with a relative dearth of points near
to this axis (see Kaiser 1987).  Such contrasting distributions are
evident between the E,S0,S0/a and disturbed galaxies in the upper and
lower panels of the figure.  While the E,S0,S0/a galaxies have a
distribution characteristic of a virially-relaxed population, the
distribution of disturbed galaxies, most especially of the disturbed
HII ELGs, is more spread out towards the caustics, thus reinforcing
the conclusion that the disturbed galaxies are a cluster infall
population.

It has long been known that spiral galaxies in clusters generally have
a higher velocity dispersion than the early-type cluster population
(e.g. Moss \& Dickens 1977; Sodr\'e et al. 1989).
Subsequent work established that it was the emission-line galaxy
population of clusters which has the higher velocity dispersion
(Biviano et al. 1997).  The present work refines these observational insights:
the infall population of the cluster comprises galaxies with {\it
disturbed morphologies}. The velocity dispersion for this population
is higher than that for the cluster and in accord with the
expected value for an infall population.  A subset of
this population (and a subset of the entire ELG population) are HII
ELGs with disturbed morphologies.  As will be discussed below (see section 
\ref{egg}), these ELGs are likely to be in the final
stages of first infall, with the highest velocity dispersion of any
cluster galaxy sample.

\subsection{Cluster halo population}
\label{chp}

\begin{figure}
\centering
\includegraphics[width=0.613\textwidth,angle=-90,bb=83 231 484 545]{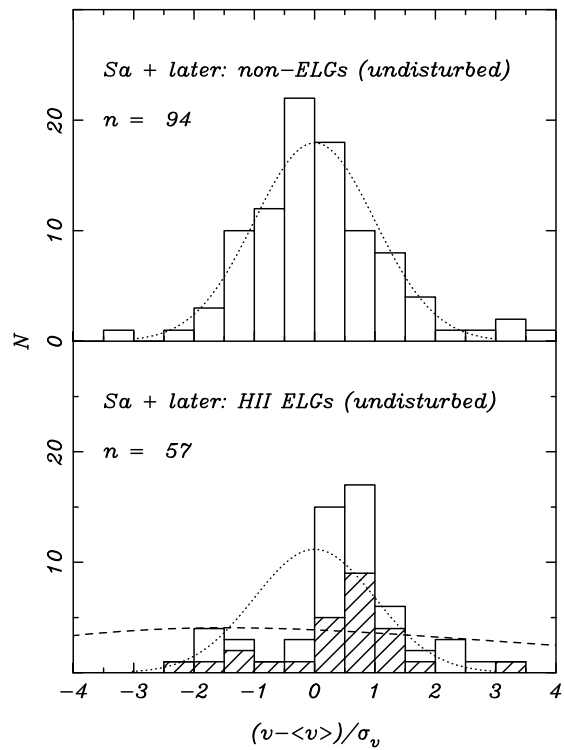}
\caption{\label{vdud} As Figure \ref{vde} for cluster ($r \le r_{vir}$) 
Sa + later galaxies with undisturbed morphology: 
(upper panel) non-ELGs and (lower panel)
HII ELGs. The hatched area of lower histogram shows the distribution
of galaxies with compact emission. The dashed line in the lower panel shows
the normalised distribution for an homogeneously distributed field population.}
\end{figure}

\begin{figure}
\centering
\includegraphics[width=0.35\textwidth,angle=-90,bb=61 92 536 657]{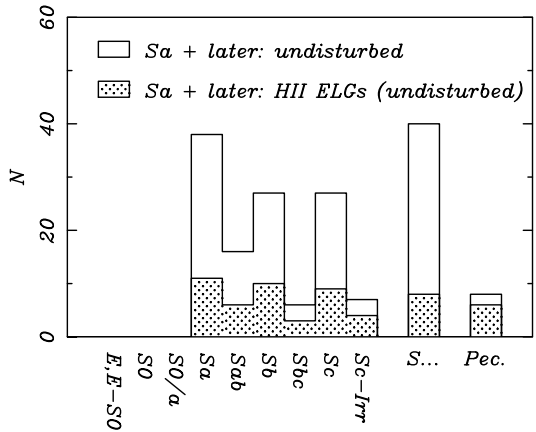}
\caption{\label{ttud} Distributions of galaxy types for Sa + later galaxies
with an undisturbed morphology: (main histogram) all galaxies, and
(shaded histogram) HII ELGs.}
\end{figure}

\begin{figure}
\centering
\includegraphics[width=0.368\textwidth,angle=-90,bb=170 230 414 533]
{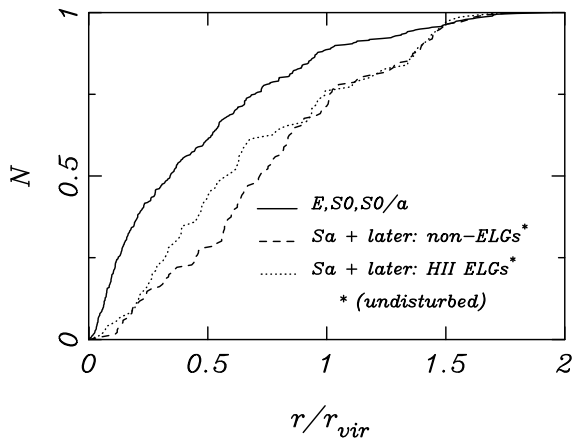}
\caption{\label{cdud} Cumulative distributions with $r_{vir}$ for 
galaxy types E,S0,S0/a (solid line, $n = 377$) and Sa + later galaxies
with an undisturbed morphology: non-ELGs (dashed line, $n = 136$) and
HII ELGs (dotted line, $n = 75$) in the ensemble cluster.  All samples
exclude galaxies with $\left| v - \bar{v} \right| > 4\sigma$.}
\end{figure}

\begin{figure}
\centering
\includegraphics[width=0.613\textwidth,angle=-90,bb=83 231 484 545]{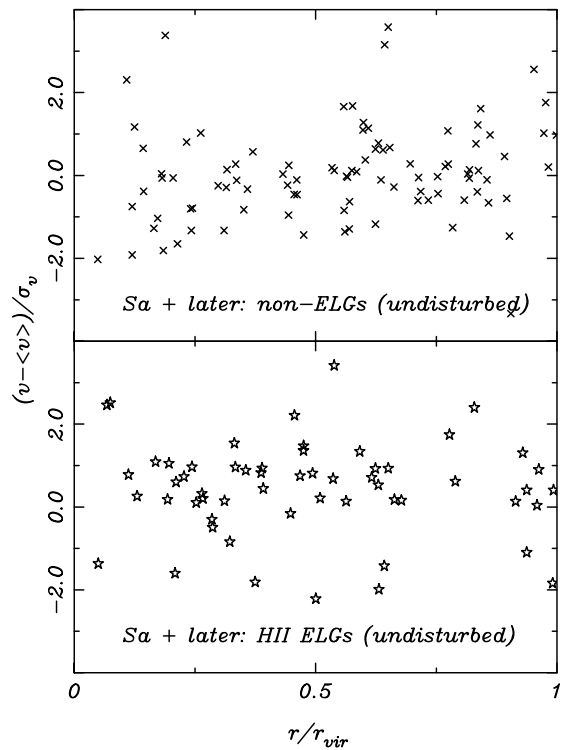}
\caption{\label{vrdud} Distribution in $r$--$v$ plane of Sa + later
galaxies with an undisturbed morphology: (upper panel) all galaxies, and
(lower panel) HII ELGs.}
\end{figure}

In Figure \ref{vdud} is shown the velocity distribution for galaxies
of types Sa + later with an undisturbed morphology.  This sample has
been divided into two sub-samples, viz. (upper panel) non-ELGs and
(lower panel) HII ELGs. There is a striking difference between the
velocity distributions for the two samples. The non-ELG population 
closely approximates a Gaussian, whereas the HII ELGs have an
asymmetrical distribution.  In this section the non-ELG population is
discussed.  Consideration of the HII ELGs is given in the next section
(\ref{pfg}).

The velocity distribution of the non-ELG population is well fitted by
a normalised Gaussian (K--S test, significance level = 0.31) with a
mean and dispersion (biweight estimators) of $-0.05 \pm 0.11$ and 1.05
respectively in good agreement with values for the E,S0,S0/a
population.

The cumulative spatial distribution with $r_{vir}$ for this population
is compared with that for the E,S0,S0/a population in Figure
\ref{cdud}.  As is seen, the distribution for the non-ELGs is less
concentrated than the corresponding early-type galaxies (K--S test,
significance level = $3 \times 10^{-12}$). The distribution of the
the non-ELGs in the $r$--$v$ plane is shown in Figure
\ref{vrdud} (upper panel).  This distribution lacks any signature of
an infall pattern in contrast to that for disturbed galaxies (cf.
Figure \ref{vrdd}).

The less concentrated spatial distribution of the undisturbed non-ELGs
of types Sa + later compared to the early-type cluster population,
and their similar Gaussian velocity distribution to the latter,
suggest that these non-emission spirals predominantly form a cluster
halo population which may be already partially or mainly virialised.

\subsection{Outer cluster population: accretion of groups}
\label{pfg}

\begin{table}
\centering
  \caption{\label{elgvd} Normalised (biweight estimator) mean
  velocities and velocity distributions for HII ELGs with undisturbed
  morphologies in the individual clusters. Samples are restricted to
  $\left| v - \bar{v} \right| \le 4 \sigma$.}
\vspace{\baselineskip}
\begin{tabular}[h]{lccccc} \hline
\multicolumn{1}{l}{\rule[-2mm]{0mm}{4mm} \hspace*{-0.5em} Cluster} &
\multicolumn{1}{c}{$\bar{v}$}  &
\multicolumn{1}{c}{$N^{+}$/$N^{-}$} \\ \hline
Abell  262 & $0.82 \pm 0.43$ & 6/1 \\ 
Abell  347 & $0.27 \pm 0.25$ & 5/0 \\
Abell  400 & $1.55 \pm 0.60$ & 2/1 \\
Abell  426 & $0.92 \pm 0.41$ & 8/2 \\
Abell  569 & $0.52 \pm 0.23$ & 7/2 \\
Abell  779 &                 & 0/0 \\
Abell 1367 & $0.41 \pm 0.22$ & 6/2 \\
Abell 1656 & $0.56 \pm 0.31$ & 11/4 \\ \hline
\end{tabular}
\end{table}

In Figure \ref{vdud} (lower panel) the velocity distribution is shown
of HII ELGs of types Sa + later, with an undisturbed morphology. As is
seen, the velocity distribution has a marked asymmetry with respect to
a normalised Gaussian with mean and dispersion (0,1) equal to that for
the early-type population (biweight estimator, $\bar{v} = 0.60 \pm
0.14$; K--S test for Gaussian fit, significance level = $3 \times
10^{-5}$). The velocity dispersion for the HII ELGs (biweight
estimator, $\sigma = 1.04$) is in agreement with that for the early-type
population.

The cumulative distribution of these ELGs with $r_{vir}$ is shown in
Figure \ref{cdud}.  In contrast to the population of HII ELGs with
disturbed morphology which follow the distribution of the early-type
population (see above, section \ref{cip}), the present ELGs have a less
concentrated spatial distribution (K--S test, significance level =
$1.6 \times 10^{-4}$).

As noted above, the undisturbed population of Sa + later galaxies may
be expected to be in the outer cluster regions or in the field. 
Can the observed asymmetry of the velocity distribution of the HII ELG
subset of this population be explained by field galaxy contamination?
In Figure \ref{vdud} (lower panel) the dashed line shows the expected
normalised velocity distribution for these ELGs, assuming they are
distributed as an homogeneous field population. This distribution was
obtained by assuming a Schechter luminosity function typical for field
galaxies with H$\alpha$ equivalent width, $W_{\lambda} \ge 20 {\rm
\AA} $ ($M^{*} = -19.17$, $\alpha$ = -1.24, see  Madgwick et
al. 2002).  Note that the form of the distribution is insensitive to
the exact value of the faint-end slope, since galaxies at the survey
limit have an absolute magnitude approximately equal to the Schechter
characteristic magnitude at the cluster distance.  The expected
distribution for an homogeneous field population shows that a
significant fraction of the actual population of ELGs is more
clustered, and is likely to be associated with the outer cluster
regions rather than with the field.  It also shows that the asymmetry
in the velocity distribution for these ELGs cannot be ascribed to
contamination by field galaxies, since such contamination would lead
to a {\it negative} asymmetry, rather than the positive one
observed.

\begin{figure}
\centering
\includegraphics[width=0.350\textwidth,angle=-90,bb=76 96 522 657]{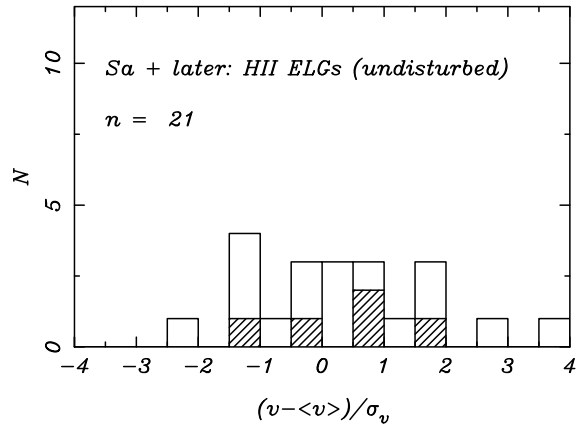}
\caption{\label{udef} As Figure \ref{vde} for Sa + later galaxies
in the supercluster field ($r > r_{vir}$) with undisturbed morphology
which are HII ELGs. The hatched area of the histogram shows the 
distribution of galaxies with compact emission.}
\end{figure}

This conclusion is reinforced by consideration of the velocity
distribution of HII ELGs with undisturbed morphology of types Sa +
later which lie beyond one virial radius from the cluster centre in
the supercluster field shown in Figure \ref{udef}.
As is seen, there is an absence of the asymmetry shown in the distribution
of corresponding cluster galaxies, and which might be expected if this
asymmetry was due to field galaxy contamination.

Assuming that a significant fraction of this ELG sample is associated with
the outer regions of the clusters, and further assuming that
individual galaxies have a randomly distributed isotropic accretion onto
the clusters, the probability of obtaining the observed velocity
distribution (i.e. 79\% of galaxies with $v >
\bar{v}$; $n = 57$) is $P \sim 5 \times 10^{-6}$. 
This asymmetry in the velocity distribution is not due to the effect
of a few clusters in the sample.  In Table \ref{elgvd}, normalised mean
velocities and values of $N^{+}$, $N^{-}$ the numbers of galaxies
with velocities greater or less than the cluster mean respectively,
are listed for individual clusters. As is seen, in all 7 cases,
$N^{+} > N^{-}$.

However the assumption that individual galaxies have a randomly
distributed isotropic accretion onto the clusters may be
questioned. Clusters of galaxies are expected to form at the
intersections of filamentary structures, and we might accordingly
suppose that accretion takes place preferentially along these
filaments. Moreover galaxies are likely to accrete onto clusters in
large agglomorations which have already condensed out of the general
field.  In which case, there may be a predominant bulk inflow
associated with a cluster, or group of clusters. For the sample of 8
clusters which comprise the ensemble cluster, one cluster (Abell 779)
has no galaxies in the present sample; two clusters (Abell 1367, 1656)
form a double cluster with separation $\sim$ 33 Mpc; and three further
clusters (Abell 262, 347, 426) form a group of clusters with mean
separation $\sim$ 18 Mpc. Accordingly if there are predominant bulk
inflows associated with the double cluster, the group of clusters and
individually with the remaining clusters (Abell 400, 569), it is no
longer so surprising that these may produce the type of asymmetrical
velocity distribution found for the present galaxy sample.

Although further investigation is needed, it is provisionally
concluded that the sample of HII ELGs with undisturbed morphology of
types Sa and later are likely to comprise a population in process of
accretion onto the cluster, but which have not yet undergone 
relaxation within the inner regions of the cluster ($r < r_{vir}$)
where their bulk streaming velocities would be randomised (cf.
section \ref{dgin}).  This view
is consistent with the lack of a high velocity dispersion for this
sample ($\sigma_{v} = 1.04$), their undisturbed morphology, their
comparatively high, often disc-wide, star formation, and the lack of
any infall signature in their $r$--$v$ distribution (see lower panel
of Figure \ref{vrdud}).

\subsection{Galaxy--galaxy interactions and mergers}
\label{ggm}

\begin{table}
\centering
  \caption{\label{dgals} Galaxies with companions.  All cluster galaxy
populations are restricted to types Sa + later, and exclude galaxies
with $\left| v - \bar{v} \right| > 4\sigma$.}
\vspace{\baselineskip}
\begin{tabular}[h]{lccc} \hline
%\multicolumn{1}{l}{\rule[-2mm]{0mm}{4mm} \hspace*{-0.5em} 
%Cluster population} & 
\multicolumn{1}{l}{Cluster population} & 
\multicolumn{1}{c}{$n$} &
\multicolumn{1}{c}{With}  &
\multicolumn{1}{c}{With}  \\
\multicolumn{1}{l}{($r \le r_{vir}$)} &&
\multicolumn{1}{c}{companions} &
\multicolumn{1}{c}{companions} \\
 &&&
\multicolumn{1}{l}{or `peculiar'} \\ \hline 
{\it Undisturbed galaxies} &&& \\ 
\hspace{1em} All       & 134 &  7.5\% & 12.7\%  \\ 
\hspace{1em} HII ELGs  &  47 &  6.4\% & 17.0\%  \\
{\it Disturbed galaxies} &&& \\
\hspace{1em} All        & 30 & 53.3\% & 73.3\%  \\
\hspace{1em} HII ELGs   & 16 & 37.5\% & 75.0\%  \\ \hline
\end{tabular}
\end{table}

\begin{table}
\centering
  \caption{\label{dgcc} Disturbed galaxies of types Sa + later within
the ensemble cluster ($r \le r_{vir}$).  Galaxies with 
$\left| v - \bar{v} \right| > 4\sigma$ have been excluded.}
\vspace{\baselineskip}
\begin{tabular}[h]{@{}l@{}c@{}l@{}c@{}c@{}c@{}} \hline
\multicolumn{1}{@{}l}{CGCG} & 
\multicolumn{1}{@{}c}{Comp.} &
\multicolumn{1}{@{}l}{Type} & 
\multicolumn{3}{c}{Companions} \\ \cline{4-6}
&&&\multicolumn{1}{@{}c}{$L_{H2}$/$L_{H1}$} &
\multicolumn{1}{@{}c}{$\Delta x$} &
\multicolumn{1}{@{}c}{$\left| \Delta v \right|$} \\
 &&&&
\multicolumn{1}{c}{(arcsec)} &
\multicolumn{1}{c}{(km ${\rm s}^{-1}$)}  \\
\multicolumn{1}{@{}c}{(1)} &
\multicolumn{1}{@{}c}{(2)} &
\multicolumn{1}{@{}c}{(3)} &
\multicolumn{1}{@{}c}{(4)} &
\multicolumn{1}{@{}c}{(5)} &
\multicolumn{1}{@{}c}{(6)} \\ \hline
 522-005  *  & C     & pec          & \ldots &  18 & \ldots \\
 522-063  *  & (C:)  & S--Irr       &&& \\
 538-043  *  &       & pec          &&& \\
 538-048  *  & C:    & S pec        & 0.9 & 134 & \ldots \\
 538-056  *  & C     & pec          & 6.3 &  34 & 427 \\
 539-025     & (C:)  & SB pec       & 0.4 & 207 & \ldots \\
 539-029     & (C:)  & S            & 0.1 &  86 & \ldots \\
             &       &              & 1.5 & 172 & 244 \\
 415-042     & (C:)  & S: pec       & 1.0 &  87 &  13 \\
 540-036     &       & S:c: pec     &&& \\
 540-057     &       & S?           &&& \\
 540-064     & C:    & SBb          & 0.3 & 112 & \ldots \\
 540-090     & C:    & pec          & 0.1 &  50 & \ldots \\
             &       &              & 0.1 &  49 & \ldots \\
 540-093     & (C:)  & Sb           & 0.9 & 110 &  79 \\
 540-112A *  & C     & pec          & \ldots &  10 & 133  \\
 541-011     & (C:)  & SB:b: pec    & \ldots & 139 & \ldots \\
 541-017     &       & pec          &&& \\
 234-071     & (C::) & SB: pec      & 0.3 & 160 & \ldots \\
             &       &              & 0.5 & 183 & 242 \\
 234-079A    & C     & S: pec       & 1.4 &  23 &  38  \\
 234-079B    & C     & S: pec       & 0.7 &  23 &  38 \\
 181-023  *  & C     & S            &&& \\
  97-062     &      & Sa: pec      &&& \\
  97-063     &      & S pec        &&& \\
  97-068     & (C:) & SBc pec       & 0.4 & 158 &   2 \\
  97-079  *  & C    & S? pec       &&& \\
  97-087  *  & C     & Sd pec       & 4.5 & 200 & 409 \\
  97-093  *  & C     & Sa:          &&& \\
  97-102A    & C    & Sa            & 0.9 &  25 &   4 \\
  97-122     &      & Sb pec       &&& \\
  97-125  *  &       & Sa: (pec)    & 0.5 & 101 &  22 \\
 127-046  *  & (C:) & S(B)bc pec   &&& \\
 160-064  *  &      & pec          &&& \\
 160-075  *  &      & pec          &&& \\
 160-140  *  & C    & S             & \ldots &  35 & 374 \\
 160-148A    & C    & S: pec        & 2.7 &  21 & 207 \\
 160-148B    & C    & S: pec        & 0.4 &  21 & 207 \\
 160-173     &      & S            &&& \\
 160-179     &      & S: pec       &&& \\
 160-180  *  &      & pec          &&& \\
 160-191  *  &      & pec          &&& \\ \hline
\end{tabular}

\begin{flushleft}
Notes on individual objects:

522-005: companion in common envelope with galaxy; likely on-going merger.

522-063: = UGC 1387. UGC note: ``Chaotic fragments of spiral
arms. May be disturbed by companion''. There is a companion at $\Delta x
= 109$ arcsec; however the velocity difference with the galaxy
($\Delta v = 765$ km ${\rm s}^{-1}$) is rather high for a slow encounter.

538-043: R, H$\alpha$ imaging shows that this is an interacting pair of
galaxies in a common envelope.

538.048: the structure of this highly distorted galaxy in the R band
is suggestive of an on-going merger. Accordingly the disturbance is more
likely to be caused by this merger than by interaction with the listed
companion.

538-056: ``wispy ring [galaxy] with an elliptical companion located near
the minor axis'' (Horellou et al. 1995).

\end{flushleft}
\end{table}

\begin{table}
%\addtocounter{table}{-1}
%\caption{continued.}
\begin{flushleft}

540-112A: this is the north component of an interacting double system,
cf. Meusinger, Brunzendorf \& Krieg 2000.

181-023: The galaxies NGC 2831/2832 = Arp 315 are close companions,
but the velocity difference ($\Delta v \sim 900$ km ${\rm s}^{-1}$)
between these and CGCG 181-023 = NGC 2830 is rather high for a slow
encounter.

97-079: CGCG note: double galaxy (Zwicky et al. 1960--68). The double
structure is clearly visible in J,K and H$\alpha$ imaging.

97-087: Gavazzi et al. (2001) have undertaken a detailed study of the
2D velocity field for this edge-on galaxy. These authors conclude that
the complex kinematical behaviour, as well as a double nucleus can
best be explained as two superposed interacting galaxies.  This is the
most likely explanation for the disturbance of this galaxy, rather than
interaction with the listed companion in the Table.

97-093: There is a companion at $\Delta = 39$ arcsec, but the velocity
difference ($\Delta v = 714$ km ${\rm s}^{-1}$) is rather large for
a slow encounter.
 
97-125: This galaxy was not classed as having a companion due to its large
velocity difference with the presumed possible companion NGC 3860 ($\Delta v
= 2676$ km ${\rm s}^{-1}$; see Paper IV).  However the galaxy is interacting
with CGCG 97-114, cf. Sakai et al. (2002).

127-046: this galaxy appears as a double galaxy in R, H$\alpha$ imaging.
The double structure is even more clearly visible in J,K.

160-064: this galaxy is seen to have a double structure in H$\alpha$,
J and K imaging.

160-075: This galaxy is close to NGC 4860 ($\Delta x = 37$ arcsec) but the
velocity difference between the two galaxies ($\Delta v = 1456$ km 
${\rm s}^{-1}$) is too great for a slow encounter.

160-140: ``A striking emission is detected southwest of this
galaxy...which suggests a close interaction with
its neighbor DRCG 27-62.'' (Bravo-Alfaro 2000).
 
160-180: J,K imaging shows that this is a double galaxy with the main 
galaxy and smaller companion in a common envelope.  The double structure
is very evident in H$\alpha$ but only faintly visible in R.

160-191: ``UCM1304+2907 MCG+05-31-133, the interactive fragmented system
VV 841 and KUG1304+291. Irr Coma system with an embedded disc aligned
near NS direction'' (Vitores et al. 1996).

\vspace{\baselineskip}
Explanation of the columns of the Table:

Column 1. CGCG number (Zwicky et al. 1960--68). The numbering of CGCG
galaxies in field 160 (Abell 1656), which has a subfield covering the
dense central region of the cluster, follows that of the listing of
the CGCG in the SIMBAD data base.  The enumeration is in strict order
of increasing Right Ascension.  An asterisk following the CGCG number
indicates a note on the galaxy below the Table.

Column 2.  Original companion parameter as listed in Papers III and
IV. Note that parentheses indicate that the probability of the companion
being a chance superposition, $P > 0.05$.

Column 3. Galaxy type taken from Papers III and IV.

Columns 4--6: Possible companions. Companions with a velocity
difference with the galaxy, $\left| \Delta v \right| > 500$ km 
${\rm s}^{-1}$, have been omitted from the Table.

Column 4: Ratio of H-band luminosities of companion and galaxy.
This ratio corresponds to the approximate dynamical mass ratio 
(cf. Gavazzi, Pierini \& Boselli 1996). Values for the individual
luminosities were taken from NED.

Column 5: Angular separation of galaxy and companion.

Column 6: Velocity difference of galaxy and companion.

\end{flushleft}
\end{table}

In Paper IV, galaxies of types Sa + later were ranked according to the
presence of an apparent nearby interacting companion ($>$ 20\% of the
size of the galaxy, and within $\sim$ 5 galaxy diameters). If
velocities were available for both the galaxy and companion, and
$\left| \Delta v \right| > 1500$ km ${\rm s}^{-1}$, the galaxies were
no longer considered to have `real' companions. (Either the projected
companion is a chance superposition, or $\left| \Delta v \right|$ is
too large for a slow gravitational encounter.) Likely projected
companions were also rejected if the probability of a chance
superposition, $P > 0.05$, based on estimates of the local surface
density.  Remaining galaxies with ranks C, C: (see Paper IV) were
considered to have `real' companions.  In Table \ref{dgals} the
percentages of galaxies of types Sa + later with `real' companions are
listed for several cluster galaxy populations, viz. undisturbed
galaxies; undisturbed galaxies with HII emission; disturbed galaxies;
and disturbed galaxies with HII emission.  

As is seen, both populations of disturbed galaxies have a much higher
($\sim$ 40--50\%) fraction of galaxies with companions than the
corresponding fraction ($\sim$ 7\%) for the populations of undisturbed
galaxies. For both populations of disturbed galaxies, a $\chi^{2}$
test gives the significance level for the observed fraction of
galaxies with companions as compared to the remainder of the Sa +
later population, as follows: disturbed galaxies, $P = 5 \times
10^{-10}$; and disturbed galaxies with HII emission, $P = 2.1
\times 10^{-3}$.

In previous work (see Papers IV and V) it was shown that galaxies
classified as peculiar\footnote{Note that galaxies were classified
as `peculiar' if their morphology was such that it could not be
assigned a Hubble type; in contrast, galaxies of all types were noted
as disturbed if their stellar distributions showed significant asymmetry
or irregularity. For a more detailed description, see Paper IV.}
show no tendency to have tidal companions,
although a very high percentage ($\sim$ 76 per cent) of these galaxies
show compact emission, which is otherwise associated either with a
bar, or with circumnuclear starburst emission caused by galaxy--galaxy
interactions. It was noted that a natural explanation for this result
is that peculiar galaxies are predominantly on-going mergers in which
the companion is already indistinguishable from its merger partner,
and the compact emission arises from the starburst induced by the
merger.

Accordingly, peculiar galaxies may also be considered to have `real'
companions, although in this case the companion has already begun
merging with the galaxy.  In Table \ref{dgals} are listed the
fractions of galaxies of the cluster galaxy populations which are {\it
either} classified as peculiar, {\it or} which have a distinct visible
companion.  The fractions of both populations of disturbed galaxies
with such mergers or companions is seen to be very high, and
significantly larger than for the remaining Sa + later population,
viz. disturbed galaxies, 73\% ($\chi^{2}$ test, $P = 2 \times
10^{-12}$); and disturbed galaxies with HII emission, 75\% ($\chi^{2}$
test, $P = 1.7 \times 10^{-5}$).

These results suggest that, for the cluster disturbed galaxies, the
cause of the disturbance is predominantly galaxy--galaxy slow
gravitational encounters and interactions.  In order to test this
conclusion further, each galaxy in the sample of cluster disturbed
galaxies has been examined in more detail.  The results of this
examination are given in Table \ref{dgcc}.

The Table lists all 39 galaxies in the cluster ($r \le r_{vir}$;
$\left| v - \bar{v} \right| > 4\sigma$) disturbed galaxy sample with
types Sa + later, together with possible companions. The CGCG
identification for each galaxy is given in column (1). In column (2)
are listed the original companion ranking for each galaxy taken from
Papers III and IV. Note that if the ranking is given in parentheses,
this means that all of the identified possible companions have a
probability of being a chance superposition, $P > 0.05$ (see Paper
IV). In column (3) galaxy types, again taken from Papers III and IV,
are listed.  Columns (4)--(6) of the Table give information on
possible companions: the ratio of H-band luminosities of companion and
galaxy, as indicative of the mass ratio (cf. Gavazzi et al. 1996); the
angular separation from the galaxy; and (where known) the absolute
velocity difference, $\left| \Delta v \right|$, with the galaxy. Since
the main interest is to identify slow encounters, all possible
companions with $\left|
\Delta v \right| > 500$ km ${\rm s}^{-1}$ were excluded from the Table.

Detailed notes on a number of the galaxies are given below the
Table. For a subset of the galaxies, R and narrow-band H$\alpha$
imaging from the JKT and Nordic Optical Telescope on La Palma, and J,K
imaging from UKIRT are available; these imaging data are being used
for future studies of the survey galaxies (Thomas et al.  2006; Moss
et al. 2006). From these images, evidence was found for double
structure and/or interaction with a companion for a number of the
galaxies (viz. CGCG 538-043, 538-048, 97-079, 127-046, 160-064
and 160-180; see notes below the Table).

From the data in Table \ref{dgcc}, some 7 galaxies can be considered
as `confirmed' members of galaxy-galaxy interactions.  Firstly there
are galaxies with close companions which have a small velocity
difference, $\left| \Delta v \right|$, together with a low probability
of the companions being chance superpositions. (viz. CGCG nos.
540-112A, 97-102A, 234-079A and B, and 160-148A and B).  In addition,
CGCG 97-125 is in a well-studied interacting system (cf. Sakai et
al. 2002).

There are an additional 11 galaxies which are probable components of
interacting systems.  These include galaxies with a double
structure (viz. CGCG 522-005, 538-043, 538-048, 97-079, 127-046,
160-064 and 160-180); and galaxies which individual studies have shown
to be probable members of an interacting system (viz. CGCG 97-087,
538-056, 160-140 and 160-191).  Interestingly, four of the galaxies
which have been revealed to have double structure by deeper imaging
(viz. CGCG 522-005, 538-043, 160-064 and 160-180) are typed as
peculiar, which gives support to the suggestion that peculiar galaxies
are likely to be on-going mergers.

The combined number of galaxies with `confirmed' and probable
interactions and/or mergers is thus 46\% of the disturbed galaxy sample. In
addition, there are a further 7 galaxies which are possible members of
interacting systems.  These include galaxies with a close companion
which has a low probability ($P < 0.05$) of being a projected
companion, although the velocity difference with the galaxy is not
known (viz. CGCG 540-064 and 540-090); and galaxies with a companion
which, while it has a significant probability ($P > 0.05$) of being a
projected companion, its velocity difference with the galaxy is small
(viz. CGCG 539-029, 415-042, 540-093, 234-071 and 97-068).  Further
galaxies which may be included as possible merger systems are those
typed as peculiar but for which deeper imaging is not yet available
(viz. CGCG 541-017 and 160-075). Including possible interactions and
mergers, the combined percentage of the cluster disturbed galaxy
sample which may belong to interacting/merging systems is thus
69\%. It is to be noted that this percentage is in fact a lower limit,
since further detailed study of individual objects may reveal 
additional interacting systems.

It is thus concluded that for $\sim$ 50\%--70\% or greater of the
cluster disturbed galaxy population, their moderate or severe 
disturbance is likely to be due to galaxy--galaxy interactions
associated with slow encounters.  
 
\section{Discussion}

\subsection{Disturbed galaxies as the infall population}
\label{dgin}

As noted in section \ref{cip}, the infall population for the ensemble
cluster is identified as (predominantly) comprising those cluster
galaxies which have moderate or severe tidal disturbance of their
stellar populations.  This identification was made on the basis that
these galaxies, which are mainly of type Sa and later (71\% of the
sample), have a higher velocity dispersion ($\sigma = 1.42$) than all
remaining cluster galaxies, with a value in good agreement with that
expected for an infall population. Remaining cluster galaxies divided
into samples of undisturbed galaxies of types Sa + later, with and
without emission, have velocity dispersions consistent with that for
the cluster early-type galaxies ($\sigma = 1.02$).  Confirmation that
the disturbed galaxy population is an infall population is provided by
the relative distributions of early-type and disturbed galaxies in the
$r$--$v$ plane (see section \ref{cip}).

The inner cluster region ($r \le 0.4r_{vir}$) shows a sharp increase
in the surface (and expected space) density of galaxies of types Sa +
later, as compared to the lower, roughly uniform, surface density for
$r > 0.4r_{vir}$. Together with this change of surface density, there
are notable changes in galaxy disturbance. The fraction of Sa + later
galaxies which are disturbed doubles within the inner cluster region
to $\sim 30\%$ as compared to $\sim 15\%$ for $r > 0.4r_{vir}$.
Moreover there is a 50\% increase in the fraction of disturbed Sa +
later galaxies which have HII emission from $\sim 40\%$ for $r >
0.4r_{vir}$ to $\sim 60\%$ for $r\le 0.4r_{vir}$.

The galaxies which are included in the present survey are the
relatively sparse population of giant ($M_{B} \le -18.5$) galaxies.
For these cluster galaxies the relaxation times are quite short.  The
two-body relaxation time, $t_{r}$, may be roughly estimated as,

\vspace{\baselineskip}
 
$t_{r} \sim \frac{\displaystyle 0.06 N}{\displaystyle {\rm
ln}(0.15 N)} \times t_{d}$

\vspace{\baselineskip}

\noindent where $N$ is
the number of cluster galaxies and $t_{d}$ is the cluster crossing
time (cf. Conselice, Gallagher \& Wyse 2001).  For the individual
clusters in the survey, $t_{r} \sim t_{d}$ especially within the high
density core region ($r \le 0.4 r_{vir}$) of the cluster.  Since the
infall population is not virialised, this suggests it is on 
first infall or, at the least, is a relatively recent arrival in the
cluster.  Furthermore the enhanced starbursts associated with
this population tend to confirm this, since the disturbed galaxies
evidently still retain substantial gas which may be expected to be
stripped by tidal, ram-pressure and harassment effects by more
prolonged exposure to the cluster environment. 

\subsection{Mergers and galaxy harassment}

\begin{figure}
\centering
\includegraphics[width=0.368\textwidth,angle=-90,bb=66 71 540 676]{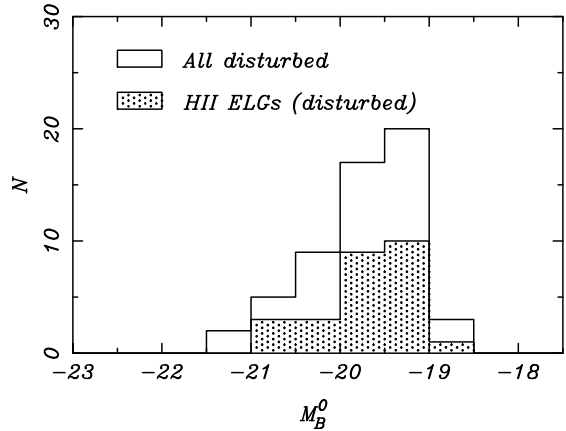}
\caption{\label{amd} Distributions of absolute magnitudes for all galaxies 
with a disturbed morphology and (shaded histogram) HII ELGs with a
disturbed morphology. }
\end{figure}

Many authors have attributed the transformation of cluster spirals to
S0s to the effects of galaxy harassment, i.e. frequent high-speed
galaxy encounters within the cluster.  Notwithstanding the strong
evidence for galaxy interactions and mergers (see section \ref{ggm}
above), can this mechanism explain the the tidal disruption and
associated circumnuclear starburst emission of the infalling galaxy
population?

Originally, galaxy harassment was proposed as a mechanism to explain
the origin of dwarf elliptical galaxies in clusters as remnants of
harassed low-luminosity (e.g. $L^{*}/5$ and $L^{*}/20$) bulgeless disc
galaxies (Moore et al. 1996; Moore, Lake \& Katz 1998).  The effect of
multiple close ($d < 50$ kpc) fast ($\Delta v \sim 1500$ km ${\rm
s}^{-1}$) encounters with giant cluster galaxies ($L \ge L^{*}$) on
the fragile disc of such a galaxy was shown to drastically alter its
morphology to resemble that of a dwarf elliptical. While galaxy
harassment is able to describe the formation of dwarf galaxies as well
as the fueling of low luminosity AGNs and the destruction of
low-surface brightness galaxies in clusters, it is evident that this
scenario is hardly applicable to the present sample of the cluster
infall population. For a value of $M^{*} = -20.5$ ($H_{0} = 70$ km
${\rm s}^{-1}$ ${\rm Mpc}^{-1}$) adopted by Moore et al. (1998), their
most luminous model disc galaxy has an absolute magnitude, $M$ =
-18.8, which is fainter than most disturbed galaxies in the ensemble
cluster (see Figure \ref{amd}).  Furthermore most of these disturbed
galaxies are not bulgeless disc galaxies of types Sc--Sd as required
by the model (see Figure \ref {ttd}).

In further work, Moore et al. (1999) discuss the effect of galaxy
harassment on luminous ($L \sim L^{*}$) disc galaxies.  Their
simulations show that luminous disc galaxies with a significant bulge
component are stable against the effects of galaxy harassment. This
makes it difficult to induce the non-axisymmetric structures in the
discs of these galaxies which are required to provide the
gravitational torques to drive gaseous inflow to fuel a central
starburst. As Mihos (2004) has noted, it is hard to explain strong
starburst activity in luminous spirals by galaxy harassment alone;
rather fast encounters tend to trigger a modest disc-wide response of
star formation. However, as has been seen above (see Figure \ref{vdd}),
most (88\%) of the detected HII emission in the disturbed galaxies of
the infall population is classified as compact, which has been
identified in previous work as most likely due to circumnuclear
starburst activity (see Papers IV, V). These disturbed galaxies with
compact emission are found more frequently in the cluster than in the
field (see Paper V). Some 40\% of disturbed galaxies show compact HII
emission in contrast to only 17\% of undisturbed galaxies of types Sa +
later (see Figure \ref{vdud}). These results support a scenario of 
slow gravitational encounters for the disturbed galaxies rather than
that of the fast encounters associated with galaxy harassment.  (For
further discussion, see Thomas et al. 2006).

One of the motivations for the development of the theory of galaxy
harassment was the claim that, for faint disturbed disc galaxies in
intermediate redshift ($z \sim 0.5$) clusters, there was often no sign
of an interacting companion, and that therefore an alternative
explanation to such interactions was needed in order to explain the
observed disturbance (cf. Moore et al. 1996).  However, whatever may
be the case for faint disc galaxies, there has never been any doubt
that for {\it giant} cluster galaxies, interactions and on-going
mergers in these intermediate redshift clusters are abundant
(e.g. Lavery \& Henry 1988; Lavery, Pierce \& McClure 1992; Dressler
et al. 1994a, 1994b). 

Just as galaxy--galaxy interactions and mergers are common among the
giant disc galaxy population in intermediate redshift clusters, so it
has been shown above (section \ref{ggm}) that such interactions and
mergers are frequent among the non-virialised infall population of
low-redshift clusters and can provide a natural explanation for galaxy
disturbance for a high proportion ($\sim$ 50--70\%) of the members of
this population, without need of any recourse to explanations based on
fast encounters, even supposing such explanations could account for
the degree of disturbance observed. For the remaining members of this
population without an obvious cause of disturbance, neither galaxy
harassment nor undetected minor mergers can be ruled out.
Nevertheless, it can be concluded that slow, rather than fast
encounters are the predominant cause of the gravitational disruption
of giant disc galaxies in low-redshift clusters.
 
\subsection{Enhancement of galaxy--galaxy interactions and mergers in
the infall population}
\label{egg}

The results obtained above imply an {\it enhancement} \/ of
galaxy--galaxy interactions and mergers in the cluster infall
population as compared to the field.  For the field population of Sa +
later galaxies ($r > r_{vir}$; $n = 84$), some 13\% ($n = 11$) are
disturbed, and may be assumed to be in interacting or merging systems.
In contrast, as has been seen, some $\sim$ 50--70\% of the infall
population are either interacting or merging with other galaxies. This
is an enhancement by a factor of $\sim$ 4--5 compared to the field
population and is statistically very significant ($\chi^{2}$ test, $P <
3 \times 10^{-7}$).

What is the cause of this enhancement of galaxy--galaxy interactions
and mergers in the infall population?  It is generally accepted that
because of the high cluster velocity dispersion, the
interaction/merger rate in (virialised) clusters is expected to be low
(e.g. Ostriker 1980; Ghigna et al. 1998). What factors, associated
with the infall population, could cause the interaction/merger rate to
increase?

One expected difference between the infall population and the relaxed
virialised population of the cluster is the
predominance of galaxy groups in the infall population. These groups
are likely to be subsequently destroyed by the tidal field of the
cluster. Distortions of the orbits of galaxies in infalling groups by
the tidal field of the cluster may increase galaxy--galaxy
interactions (cf. Mihos 2004).  Another suggestion, due to Sato \&
Martin (2006), is that group--group encounters in the infalling
population could enhance galaxy--galaxy mergers. However it may be
questioned whether either of these mechanisms alone would be capable
of enhancing the interaction/merger rate by the large factor required.

A promising mechanism to greatly enhance the interaction/merger
rate in the infall population is gravitational shocking as proposed by
Struck (2005).  Struck notes that cold dark matter simulations and
observations are in good agreement regarding a common density profile
across a range of structures from galaxy halos to the dark halos of
large clusters. Moreover observations suggest that the central density
decreases slowly with mass in dark matter halos. Accordingly for
roughly comparable group and halo core densities, the passage of a
group through the cluster core would would substantially increase the
instantaneous group halo mass.  Since the typical core crossing time
and group free-fall time are comparable, there is time for group
galaxies to be pulled into a much denser and compact configuration.
In the case that group and cluster core halo densities are roughly
comparable, galaxies could be pulled in to roughly half their distance
from the (group) core, increasing the galaxy density by nearly an
order of magnitude and their collisions by a factor of 100 (density
squared). In these encounters, dynamical friction will dissipate
relative orbital energy, leading to galaxy--galaxy mergers.  This
mechanism would readily explain the enhanced galaxy interaction/merger
rate found for the infall population of the ensemble cluster.  The
gravitational shocking discussed by Struck is for groups passing
through the cluster core; clearly, infalling groups encountering any
existing cluster sub-structure may also be expected to experience some
degree of gravitational shocking. Thus infalling groups may be subject
to a series of gravitational shocking events on their infall to the 
cluster centre, each contributing to the total interaction/merger
rate.

In section \ref{cip} it was noted that disturbed HII ELGs have a
higher velocity dispersion ($\sigma_{v} = 1.75$) than for the entire
sample of disturbed galaxies ($\sigma_{v} = 1.42$). A similar effect
is evident in comparing disturbed HII ELGs of types Sa + later with
all disturbed galaxies of these types ($\sigma_{v} = 1.69,1.55$ for
the two samples respectively). Gravitational shocking may help to
explain this effect. Since not only the frequency, but also the
strength of encounters is dependent on galaxy density, the strongest
encounters leading to circumnuclear starbursts are most likely to
occur where gravitational shocking is most effective, i.e. near the
centres of clusters. Since the infalling groups which are the sites of
such mergers will be near maximum infall velocity, the resulting HII
ELGs will generally be expected to have a higher velocity dispersion
than the overall population of disturbed galaxies.

\subsection{Transformation of spirals to S0s by unequal-mass and minor mergers}

\begin{figure}
\centering
\includegraphics[width=0.368\textwidth,angle=-90,bb=66 71 540 676]{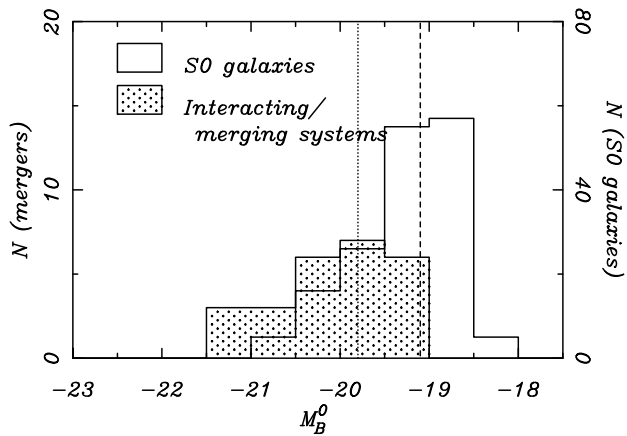}
\caption{\label{ammerg} Distribution of absolute magnitudes for interacting
and merging systems ($n=25$), and for S0 galaxies ($n=164$) in the
ensemble cluster ($r \le r_{vir}$). The interacting and merging
systems contain at least one component of type Sa + later. Survey
completion limits for the interacting and merging systems and for the
S0 galaxies are shown with the vertical dotted and dashed lines
respectively.}
\end{figure}

Theoretical studies have shown that major mergers of equal-mass
galaxies can result in the formation of an elliptical galaxy with a de
Vaucouleurs profile (e.g. Negroponte \& White 1983; Barnes 1988, 1992;
Hernquist 1992, 1993).  In contrast, mergers of unequal mass spirals
are expected to result in an S0 galaxy since the remnant retains
significant rotation and disc destruction is not complete (e.g. Bekki
1998; Bendo \& Barnes 2000; Cretton et al. 2001; Bekki et
al. 2005). Alternatively, an S0 can be formed between the merger of a
spiral and its satellite: the satellite can help to `sweep clean' the
disc of cold gas by means of a gravitationally induced bar driving gas
to the nucleus, and the galactic disc is not destroyed, but thickened
to one characteristic of an S0 (e.g. Toth \& Ostriker 1992; Quinn,
Hernquist \& Fullager 1993; Mihos et al. 1995; Walker, Mihos \&
Hernquist 1996; Mihos 2004).

Thus the interactions and subsequent galaxy mergers
identified in the Sa + later infall population of the ensemble cluster are
potentially the precursors of some fraction of the early-type cluster
population, most especially of the S0 population.  What fraction of
the present S0 population in low-redshift clusters can be accounted
for in this way?  Any such estimate is necessarily very
uncertain, but it can be attempted as follows.  In Figure
\ref{ammerg} is shown the distribution of combined absolute magnitudes for 
the components of interacting and merging systems in the ensemble
cluster (viz. the interacting and merging systems identified as
`confirmed', `probable' and `possible' in section \ref{ggm} above).
Also shown in the Figure is the absolute magnitude distribution of the
S0 population of the cluster.  The completeness limit for the S0
population is approximately, $M_{B}^{0} = -19.1$.  Following a merger,
the resulting early-type remnant is expected to fade by $\sim$ 1--2
magnitudes (cf. Larson \& Tinsley 1978). Accordingly, the six
interacting/merging systems with $M_{B}^{0} \le -20.6$ are expected to
be potential precursors of early-type galaxies brighter than the completeness
limit, $M_{B}^{0} = -19.1$. The cluster
tidal field can act to lengthen the merger time of interacting
galaxies, and may even prevent mergers in the case of very loosely bound 
interacting pairs (cf. Makino \& Hut 1997; Mihos 2004). On the other hand,
this tidal field is likely to quickly strip loosely bound tidal debris
associated with interacting galaxies, which may lead to an
underestimation of the interaction rate (see Mihos 2004).
For simplicity, these considerations will be neglected.  Thus if it is
assumed that the merger time $\sim 10^{9}$ years, then,
for a uniform infall rate into the cluster, $\sim$ 60
mergers are expected over the past 10 Gyr.  However the infall rate
is expected to be higher at earlier times (e.g. Ellingson et al. 2001).
Accordingly if it is assumed that the mean integrated infall rate is
twice the current observed rate, then $\sim$ 120 early-type galaxies
resulting from mergers are expected over the past 10 Gyr. Now the
total number of S0 galaxies with $M_{B}^{0}
\le -19.1$ in the ensemble cluster is 92.  Thus if mergers are mainly
unequal mass mergers and result predominantly in S0 galaxies, most of
the S0 population can be readily accounted for as products of 
interactions and mergers in the cluster infall population.

\section{Conclusions}

Analysis of an ensemble cluster comprising a complete
magnitude-limited sample of giant galaxies ($M_{B}^{0}$
\raisebox{-1ex}{$\stackrel{\textstyle <}{\sim}$} $-19$) in 8
low-redshift clusters has shown that disturbed galaxies in these
clusters form a infall population.  It has further been shown that the
disturbance of the stellar populations of these galaxies can readily
be explained by slow galaxy--galaxy encounters in at least 50--70\%
of cases. The resulting enhancement of slow encounters in the cluster
infall population can be attributed to gravitational shocking (Struck
2005) of infalling galaxy groups.  A simple estimate of the galaxy
merger rate demonstrates that the cluster giant S0 population can be accounted
for as the outcome of minor mergers over the past $\sim$ 10 Gyr.

Notwithstanding this result, obviously not all cluster S0 galaxies
need be products of merging galaxies in the
infall population; some early-type cluster galaxies are expected to
result from galaxy interactions in groups or the field (e.g. Zabludoff
\& Mulchaey 1998) and this may help explain the correlation between
star formation and radial distance to several virial radii from the
cluster centre (e.g. Lewis et al. 2002; Gomez et al. 2003; although
for an alternative view, see Poggianti 2004). However, contrary to
Balogh et al. (2004), it is not necessary to assume that the majority
of cluster S0s have formed by such preprocessing in the field.  Indeed
the fact that most galaxies in the infall population are spirals
(see section \ref{dgin} above), rules out such an explanation.
Rather, the bulk of transformation of spirals to S0s takes place
during, and as an inherent part of the process of virialisation of the
infall population.

Moreover, at least for the cluster galaxies, other processes may play
a significant role in the formation of cluster giant S0 galaxies.  For
the Virgo cluster, the majority of spiral galaxies have their
H$\alpha$ disks truncated most likely due to intracluster
medium-interstellar medium (ICM-ISM) stripping as a cause of the
reduced star formation (Koopmann \& Kenney 2004). A similar effect
has been found in other low-redshift clusters (Thomas et al. 2006).
Such ram-pressure stripping of gas is likely to contribute to the
transformation of cluster spirals to S0s.  Nevertheless, as the
present work has shown, it is plausible to suppose that for most
cluster giant S0 galaxies their formation has involved slow galaxy--galaxy
interactions and ensuing minor mergers.

\section*{Acknowledgements}
I would like to thank D. Carter and P.A. James for useful discussions
and an anonymous referee for helpful comments.
This research has made use of the NASA/IPAC Extragalactic Database (NED),
which is operated by the Jet Propulsion Laboratory, California Institute
of Technology, under contract with the National Aeronautics and Space
Administration.

\end{document}